\documentstyle[prd,aps,epsf,epsfig]{revtex} 

\begin{document}
\draft
\twocolumn[\hsize\textwidth\columnwidth\hsize\csname@twocolumnfalse\endcsname

%
\title{Phase transition in a spring-block model of surface fracture}
%
\author
{Kwan-tai Leung${}^\ast$ and
J\o rgen Vitting Andersen${}^\dagger$}
\address{
${}^\ast$ Institute of Physics, Academia Sinica,
Taipei, Taiwan 11529, R.O.C.\\
${}^\dagger$ Laboratoire de Physique de la
Mati\`{e}re Condens\'{e}e,
Universit\'{e}\ de Nice-Sophia Antipolis,
Parc Valrose, 06108 Nice Cedex 2, France}
\maketitle

\begin{abstract}

A simple and robust spring-block model obeying threshold dynamics
is introduced to study surface fracture of an overlayer
subject to stress induced by adhesion to a substrate.
We find a novel phase transition in the crack morphology and
fragment-size statistics when the
strain and the substrate coupling are varied.
Across the transition,
the cracks display in succession 
short-range, power-law and long-range correlations.
The study of stress release prior to cracking
yields useful information on the cracking process.

\end{abstract}

\vspace{2pc}
]


Apart from its significance in industrial applications,
cracking of a brittle material contains interesting physics
and has attracted a lot of attention recently\cite{rouxherrmann}.
Two kinds of ongoing research can be identified:
the first deals with the dynamic instability
in the propagation of a single crack\cite{crack1d},
and the second is concerned with the
collective behavior of many interacting cracks,
such as the global pattern
and distribution of fragments\cite{skjeltorp,groisman,allainlimat}.
The latter receives less attention and will be our focus here.

Although crack patterns in nature occur
in a variety of contexts\cite{walker}
over a wide range of sizes from the millimeters on
a monolayer of packed polystyrene spheres\cite{skjeltorp}
to the kilometers of giant crack networks on playas,
they are formed by the same basic processes.
Typically, when an overlayer dries and shrinks,
adhesion to a substrate resists shrinkage and induces internal stress.
The stress may be released by slipping at the interface of contact
or by cracking the overlayer;
their competition leads to the variety of crack patterns.
In recent experiments,
the scale and geometry of patterns were found to depend on
the strength of adhesion, the thickness of the layer
and boundary conditions\cite{groisman,allainlimat}.
Analytical approaches are difficult and scarce,
due to the complexity of multi-crack interactions\cite{rundleklein}.
Thus, simulations using simple models are expected
to provide useful information\cite{skjeltorp,colina,abj}.

In this Letter, we present a spring-block model of fracture
for a two-dimensional overlayer in contact with a substrate.
Our goal is to identify the major control parameters
and to explore various statistical properties of fracture
not readily accessible to experiments.
A novel phase transition is found, 
whose origin can be traced to the interplay between
stress redistributions and releases.

\underline{\sl Model:}
Our model consists of a square array of blocks,
interconnected among nearest neighbors by coil springs
with spring constant $K$ and relaxed spring length $l$.
To compare with experiments,
free boundary conditions are used, with three nearest neighbors
on edges and two at corners.
Initially, the blocks are randomly displaced about their mean
positions $\vec{r_o}=(ia, ja)$
by $(x,y)$ on a rough substrate,
where $i,j=1,\dots,L$, and $a$ is the lattice constant.
Motivated by the drying process described above,
internal stress may be introduced in various ways:
$a$ may be fixed but $l$ is decreased in time to model contraction;
or both $a$ and $l$ may be fixed to
impose an initial tensile strain ($s=(a-l)/a>0$),
while $K$ is increased to model material stiffening\cite{abj}.
As a first study, we will adopt the simpler second approach here.

Contrary to similar models of earthquakes\cite{ofc,gofc},
in which the stress arises from
relative motion of two surfaces and released by block slips alone,
the stress here is imposed uniformly by increasing $K$ 
and released by block slips {\em and\/} spring breaks
via a threshold dynamics:
a block slips to a force-free position if
the net force from its neighbors $F>F_s$ (the threshold for slipping),
and a spring breaks ($K\equiv 0$)
if the tension $b>F_c\equiv \kappa F_s$ (the threshold for cracking).
With all forces initially below thresholds,
$K$ is increased slowly with $F_s$ and $F_c$ fixed.
When either threshold is reached,
stress is dissipated and redistributed accordingly, causing
further slips and/or cracks until $F<F_s$ and $b<F_c$ everywhere.
This constitutes one {\em event\/}.
$K$ is increased only between individual events,
corresponding to slow drying where
the rate of driving is infinitesimal compared to that of relaxation.

The force exerted on a given block at $\vec{r}=\vec{r_o}+(x,y)$
by a neighboring block at $\vec{r'}$ is
given by $( |\vec{r}-\vec{r'}|-l)K$.
Since cracking is primarily due to tensile stresses,
this nonlinear dependence on the coordinates
leads to unnecessary complications in updating the configurations.
To simplify and compare with previous models, we
expand the forces to first order in $(x,y)$ to obtain
the force components on a block at $(i,j)$ (cf. \cite{gofc})
\begin{eqnarray}
F_x &=&
(a-l+x_{i+1,j}-x_{i,j})K_1 +
(x_{i,j+1}-x_{i,j})sK_2 +
\nonumber \\
& & \mbox{}
(-a+l+x_{i-1,j}-x_{i,j})K_3 +
(x_{i,j-1}-x_{i,j})sK_4
\nonumber  \\
F_y &=&
(y_{i+1,j}-y_{i,j})sK_1 +
(a-l+y_{i,j+1}-y_{i,j})K_2 +
\nonumber \\
& & \mbox{}
(y_{i-1,j}-y_{i,j})sK_3 +
(-a+l+y_{i,j-1}-y_{i,j})K_4,
\label{force}
\end{eqnarray}
where the subscript of $K$ denotes different springs.
Likewise, the tension of the spring 
between two blocks at $(i,j)$ and $(i+1,j)$ has the components
$b_x = (a-l+x_{i+1,j}-x_{i,j}) K$ and
$b_y = (y_{i+1,j}-y_{i,j}) sK$.

Without loss of generality, we hereafter
choose our units such that $a=1=F_s$.
Of the remaining parameters\cite{footnoteA1}
$\{s,\kappa,K\}$,
$K$ may be used to define the `time' $t\equiv K$.
Note that a larger $\kappa$ means
either a stronger material (larger $F_c$) or
a weaker substrate coupling (smaller $F_s$).

\begin{figure}[htp]
\epsfig{figure=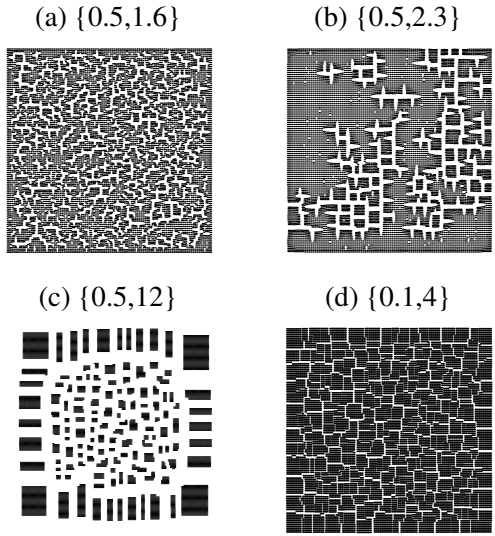,height=3.0in,angle=0} 
\caption{
Typical patterns for $L=100$. Legends stand for $\{s,\kappa\}$.
(a) Below transition, shows localized cracks and small fragments;
(b) near transition, almost percolating cracks and larger fragments;
(c) inhomogeneity well above transition;
(d) in the one-phase region, shows more isotropic cracks.
}
\label{fig1}
\end{figure}

\underline{\sl Results:}
We have simulated the model for a wide range
of values of the dimensionless parameters:
$0.01\leq s\leq 0.98$, $1\leq \kappa\leq 30$,
and system sizes $20\leq L \leq 300$.
Clearly, in the slippery limit $\kappa=\infty$,
no spring is broken and our model becomes a variant of
previous slip-stick models\cite{gofc,ofc} in their conservative limits,
with almost the same slip-size statistics.
However, notice that our model is driven multiplicatively by $K$,
whereas previous models are driven by an additive force term.
For finite $\kappa$,
the system contracts by block slips before 
it cracks into disjoint fragments.
In agreement with intuition and experiments\cite{groisman,colina},
both the waiting time $t_c$ for the first spring to break
and the mean fragment size increases with $\kappa$ (see Fig.~\ref{fig1}
and below).

\begin{figure}[htp]
\epsfig{figure=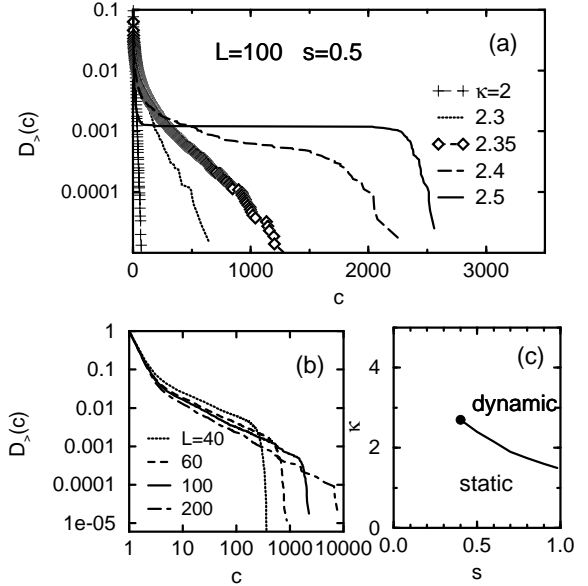,height=3.2in,angle=-0}
\caption{
(a) Cumulative distribution of crack size per event.
Note short (long)-range correlation below (above)
$\kappa^*\approx 2.4$;
(b) power-law decay $D_>(c)\sim c^{-0.63}$ at $\kappa=\kappa^*$;
(c) the phase diagram with a possible critical end point.
}
\label{fig2}
\end{figure}

Fig.~\ref{fig1} also reveals two kinds of crack patterns,
depending on an $s$-dependent ratio $\kappa^*$ for large $s$:\\
(a) {\em Static cracking\/}---For $\kappa<\kappa^*$,
cracks propagate slowly. 
Correlations in positions and orientations
are weak between successive cracks.
The cumulative distribution $D_>(c)$ of the
number of cracks per event, $c$, 
decays exponentially, 
it is short-range (see Fig.~\ref{fig2}(a)).
The spatial distribution of fragments is homogeneous,
as shown in Fig.~\ref{fig1}.\\
(b) {\em Dynamic cracking\/}---For $\kappa>\kappa^*$,
cracks percolate the sample and break it in a catastrophe.
$D_>(c)$ is long-range with a long plateau and a sharp roll-off
(see Fig.~\ref{fig2}(a)),
corresponding to a peak there in the probability density $D(c)$.
The cutoff $c_{\rm max}$ scales as $L^2$.
The distribution of fragments is inhomogeneous
deep in this region,
as shown in Fig.~\ref{fig1}(c).
Along the phase boundary at $\kappa=\kappa^*(s)$,
$D_>(c)\sim c^{-\eta}$ (see Fig.~\ref{fig2}(b))
with an exponent $\eta\approx 0.75\pm 0.05$
to $0.63\pm 0.02$ as $s$ varies from 0.8 to 0.5.

By probing the $\kappa$ and $L$ dependence of $c_{\rm max}$,
we determine the phase diagram, Fig.~\ref{fig2}(c).
The transition between static and dynamic cracking 
becomes less striking as $s$ is decreased.
Beyond the end point ($s_c,\kappa_c)$\cite{footnoteA2},
the two phases become indistinguishable.
Traversing along fixed $s$ within the one-phase region,
we find in $c_{\rm max}$ versus $\kappa$ a mild peak 
which saturates for large $L$,
and $D_>(c)$ eventually decays exponentially at large $c$,
indicating non-percolating cracks.

\begin{figure}[htp]
\epsfig{figure=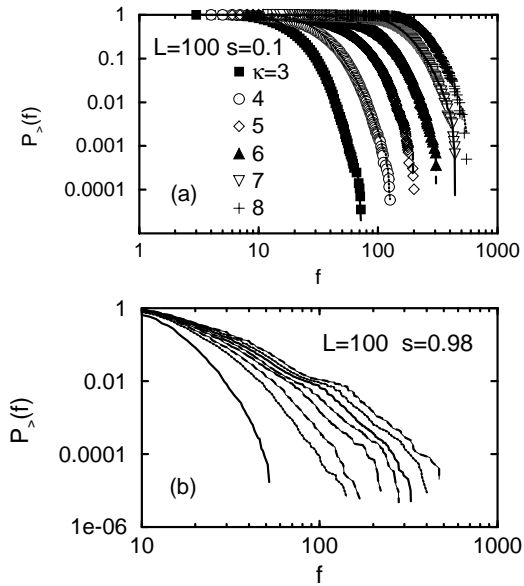,height=3.2in,angle=-0}
\caption{
Cumulative distribution of fragment area.
(a) shows good fits (lines) to log-normal distribution
for non-percolating cracking; and
(b) shows deviation from log-normal and emergence of
a kink (spatial inhomogeneity)
for dynamic cracking, where $\kappa=2$, 3, 3.5, 4, 4.5, 5, 5.5, and 6
from left to right.
}
\label{fig3}
\end{figure}

Also of particular interest is
the probability density $P(f)$ of fragment area $f$,
the center of focus of fragmentation theories\cite{redner,fragsim}
and experiments\cite{fragexpt}.
For non-percolating cracking,
$P(f)$ is best described by a log-normal distribution\cite{lognormal}
\begin{equation}
P(f)={1\over \sqrt{2\pi\sigma^2} f} \exp{-(\ln f-\mu)^2\over 2\sigma^2},
\label{lognormal}
\end{equation}
as shown in Fig.~\ref{fig3}(a)
for the cumulative distribution $P_>(f)=\int_f^{L^2} df'\, P(f')$.
On the other hand,
$P_>(f)$ for dynamic cracking deviates from log-normal
for $\kappa>\kappa^*$.
At larger $\kappa$,
a kink appears and becomes more pronounced
(see Fig.~\ref{fig3}(b)),
implying different statistics for the fragments
near the boundary and the center.
Although the tails of $P_>(f)$ suggest a power-law dependence,
we cannot be certain due to the limited range of data.

\begin{figure}[htp]
\epsfig{figure=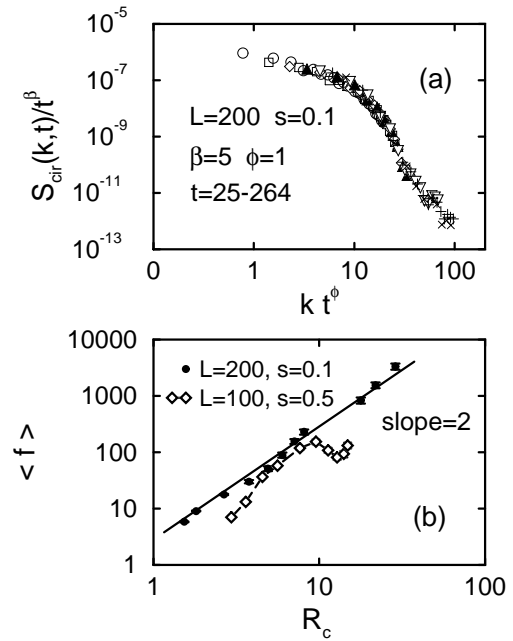,height=3.3in,angle=-0} 
\caption{
(a) Scaling of structure factor for stress field prior to cracking,
    corresponding $R(t)=30$--180.
(b) Mean fragment area scales with correlation length of stress field
    for static cracking (upper curve),
    but not for dynamic cracking (high end of lower curve).
}
\label{fig4}
\end{figure}

A consistent physical picture emerges from studying
the reduced stress field
$\varepsilon(\vec{r},t)\equiv|b|/F_c\propto
\sqrt{{\rm Tr}\mbox{\boldmath$\sigma$}^2}/F_c$,
where $\mbox{\boldmath$\sigma$}$
is the local stress tensor, and $0<\varepsilon<1$.
We compute the structure factor
$S(\vec{k},t) = \langle |\tilde \varepsilon(\vec{k},t)|^2 \rangle/L^2
-L^2 \delta_{\vec{k},0} \langle\overline{\varepsilon(\vec{r},t)}\rangle^2$,
where $\tilde \varepsilon$ is the Fourier transform,
and the overline means a spatial average.
Averaging over the circular average $S_{\rm cir}(k,t)$,
we define a characteristic length of the stress field 
$R(t)=\langle k^2 \rangle^{-1/2}_{\rm cir}$.
Before cracking occurs, block slips start from the edges
and propagate inward.
Thus, $\varepsilon$ is reduced near the edges but
enhanced linearly in time in the bulk.
These changes are reflected in the growth of $R(t)$ 
via a power law (cf. \cite{abj}) $R(t)\sim t^\phi$.
Within statistical uncertainty, $\phi\approx 1$ for all $s$
asymptotically at large $L$.
The self-similar
build-up of correlation is evident in
the temporal scaling for $t<t_c$:
$S_{\rm cir}(k,t) = t^\beta \Phi( k t^\phi)$
(see Fig.~\ref{fig4}(a)).
Hence, $R(t)$ may be interpreted either as 
the penetration depth of slip events into the bulk
or the correlation length of the stress field,
whose growth towards $L$ is a manifestation of the
approach to a critical state self-organized 
by the stick-slip mechanism\cite{ofc,gofc}.

The time $t_c$ when cracking sets in
may be deduced by subjecting 
the enhanced stress $|\vec{b}|\approx Ks$ in the bulk 
to the threshold condition
$\varepsilon \approx Ks/F_c = t s/\kappa \geq 1$.
Thus $t_c \approx \kappa /s$, which has been verified.
Whether the initial crack triggers an instability
depends on the strength of the stress $\cal S$
at a crack tip relative to the spatial fluctuation $\delta\varepsilon$.
Since $\delta b\sim KA$, 
we have $\delta\varepsilon/\varepsilon\propto A/s$.
So, for small $s$, 
$\cal S$ is suppressed relative to $\delta\varepsilon$,
fluctuations dominate and the cracks are localized.
This explains the absence of percolating cracks
in the small-$s$ regime.

For large $s$,
the transition can be understood by virtue of 
the effect of $\kappa$ on $\cal S$.
$\cal S$ is large only if there are stress transfers
from crack sides to tips 
by slipping those blocks that link to broken springs.
The ratio $F/F_s \sim Ks \approx \kappa$ determines if they slip,
where $F\sim Ks$ in the bulk (see Eq.~\ref{force})
and $K\approx t_c$ have been used.
Since $\kappa$ has little influence on $\delta\varepsilon$,
we conclude that small $\kappa$ gives rise to small $\cal S$
and hence isolated cracks.
The cracks bordering a typical fragment are 
highly localized and weakly correlated,
the usual argument leading to
a log-normal distribution applies\cite{lognormal,redner}.
On the other hand,
large $\kappa$ fulfills the slipping criterion easily,
we have substantial stress transfers at a crack opening
and a large $\cal S$,
which forces more broken springs. 
An instability is inevitable.
The strong correlations of cracks (as shown in $D_>(c)$)
invalidates the assumption for log-normal distribution.

Further insight can be gained from exploiting $R(t)$.
For non-percolating cracking,
we find the mean fragment area
$\langle f \rangle \sim R_c^2$,
where $R(t_c(\kappa))\equiv R_c(\kappa)$ 
(see Fig.~\ref{fig4}(b)).
This implies $R(t)^2$ {\em prior to\/} cracking sets a lower bound of
$\langle f \rangle$ for the {\em final\/} cracked state.
Using $t_c\approx \kappa/s$, we find
$\langle f \rangle \propto \kappa^{2\phi}$.
This simple relation highlights the crucial role of $\kappa$
in our model.
In contrast, for dynamic cracking,
this predictability is lost as
$\varepsilon$ is extensively modified by the block slips
at crack openings.

Finally, our model is robust in the following sense.
If we allow $F_c$ and $F_s$ to increase with $K$,
the simplest alternatives are:
(a) $F_c\propto K$, $F_s={\rm const}$, so $\kappa\propto K$;
(b) $F_s\propto K$, $F_c={\rm const}$, so $\kappa\propto 1/K$; and
(c) $F_s\propto F_c \propto K$, $\kappa={\rm const}$.
In the first case, the final state is entirely controlled by $F_c/K$.
It undergoes a first order transition from a crack-free
to a cracked state as $F_c/K \to s+2A\equiv \zeta$ from above.
Further below $\zeta$,
the density of broken springs quickly saturates.
The second case is trivial,
having all the springs broken at long times.
For the third case,
either nothing happens if neither thresholds are exceeded,
or the updating sequence is ambiguous if both are.
Thus, within our formalism,
our choice of rules is the only one that
evolves into a nontrivial cracked state through
the interplay of sticking, slipping and cracking,
without the need of fine tuning the parameters or initial conditions.

\underline{\sl Conclusion:}
We have introduced a robust spring-block model of
surface fracture.
Despite its simplicity,
it captures the correct trends observed in experiments and
displays rather complex behavior,
which can be understood from the evolution of the stress field
and characterized by dimensionless control parameters.
Concerning the fragment distributions,
similar transitions between log-normal and power-law distribution
have been observed in 
some recent fragmentation experiments\cite{fragexpt}.
While it is possible that the underlying mechanism
for the transition may be similar to ours,
one must be cautious because the physics with and without a substrate
could be utterly different.

The authors wish to thank Profs. Y. Brechet, C.K. Chan and D. Sornette
for illuminating discussions.
K.-t.L. is supported by
the National Science Council of ROC
under grant No. NSC85-2112-M-001-013, and
the National High Performance Computing Center of ROC;
J.V.A. wishes to acknowledge supports from
the Danish Natural Science Research Council under Grant No. 9400320,
and the European Union Human Capital and Mobility Program
contract number ERBCHBGCT920041 under the direction of Prof. E. Aifantis.



\begin{references}
\bibitem{footnoteA1}
The amplitude $A$ of the initial random displacement $(x,y)$
is an additional parameter.
We chose $A=0.02$ and found the same
qualitative features for $A=0.002$---$0.2$.
\bibitem{footnoteA2}
Larger $A$ has more disorder, giving a shorter and higher
phase boundary with a greater $s_c$.
\end{references}
\end{document}